\documentstyle[prd,aps,floats,epsfig]{revtex}

\begin{document}

\input epsf
\renewcommand{\topfraction}{1.0}
\twocolumn[\hsize\textwidth\columnwidth\hsize\csname
@twocolumnfalse\endcsname

\title{Boson stars in massive dilatonic gravity}

\author{Plamen P.~Fiziev${^1}$,  Stoytcho S.~Yazadjiev${^{1}}$}

\address{${^1}$Department of Theoretical  Physics, Faculty of Physics,
Sofia University, Bulgaria}

\author{Todor L.~Boyadjiev${^2}$ and Michail D.~Todorov${^{3}}$}

\address{${^2}$Faculty of Mathematics and Computer Science,
Sofia University, Bulgaria}

\address{${^3}$Institute of Applied Mathematics and Computer Science,
Technical University, Sofia, Bulgaria}

% \date{}

\maketitle

\begin{abstract}
We study equilibrium configurations of boson stars in the
framework of  a class scalar-tensor theories of gravity with
massive gravitational scalar (dilaton). In particular we
investigate the influence of the mass of the dilaton on the boson
star structure. We  find that the  masses of the boson stars in
presence of dilaton are close to those in general relativity and
they are sensitive to the ratio of the boson mass to the dilaton
mass within  a typical  few percent. It turns out also that the
boson star structure is mainly sensitive to the mass term of the
dilaton potential rather to the exact form of the potential .
\end{abstract}

\vskip 0.8cm

 {PACS number(s): 04.50.+h, 04.40.Dg, 95.35.+d}

\vskip2pc]
\newpage

%%%%%%%%%%%%%%%%%%%%%%%%%%%%%%%%%%%%%%%%%%%%%%%%%%%%%%%%%%%%%%%%%%%%%%%%%%%%%
\newcommand{\lfrac}[2]{{#1}/{#2}}
\newcommand{\sfrac}[2]{{\small \hbox{${\frac {#1} {#2}}$}}}
\newcommand{\ben}{\begin{eqnarray}}
\newcommand{\een}{\end{eqnarray}}
\newcommand{\la}{\label}

%%%%%%%%%%%%%%%%%%%%%%%%%%%%%%%%%%%%%%%%%%%%%%%%%%%%%%%%%%%%%%%%%%%%%%%%
\section{Introduction}

Boson stars are gravitationally bound macroscopic quantum states
made up of scalar bosons. They differ from the fermionic stars in
that they are only prevented from collapsing gravitationally by
the Heisenberg uncertainty principle. Boson stars were first
considered by Kaup \cite{K} and then by Ruffini and Bonazola
\cite{RB}. They found that the boson stars as described by
non-interacting massive, complex  scalar field had masses of the
order of ${\cal M} \approx {{\cal M}_{Pl}^2\over m_{B}}$, where
${\cal M}_{Pl}$ is the Planck mass  and $m_{B}$  is the boson
mass. In a latter work Colpi, Shapiro and  Wasserman \cite {CSW}
analyzed the consequence of switching on a quartic
self-interaction for the boson field. This results in a drastic
increase of  the mass of the boson star, which even for small
values of the coupling  constant turns out to be of the order of
the Chandrasekhar's mass when the boson mass is similar to a
proton mass. Thus, the boson stars arise as possible candidates
for non-baryonic dark matter in the universe. Although we still
have no astrophysical evidences for their existence, the boson
stars are good model to learn more about the nature of strong
gravitational fields not only in general relativity but also in
the other theories of gravity.

The most natural and promising generalizations of Einstein's
general relativity are scalar-tensor theories of gravity. The most
studied class of such theories are scalar-tensor theories with
massless gravitational scalar field including the Brans-Dicke
theory as a special case. It should be noted that an interesting
specific class of scalar-tensor theories with dilaton field arise
from the low energy limit of string theory, which confirms their
importance for current physics.

Boson stars in the framework of scalar-tensor theories with
massless gravitational scalar have been studied widely, too. The
first scalar-tensor model of a boson star was studied by Gundersen
and Jensen \cite{GJ}, who considered Brans-Dicke theory with
${\omega}_{BD} = 6.$  Their work was generalized by Torres
\cite{T}, who studied boson stars in scalar-tensor theories with
nonconstant  ${\omega}_{BD}(\Phi).$  The conclusion is that boson
stars can exist in any scalar-tensor theory of gravity with
massless gravitational scalar, as the masses of the boson stars
are always smaller than in the general  relativistic  case
irrespective of the coupling functions ${\omega}_{BD}(\Phi)$. We
should mention also the interesting paper by Z. Tao and X. Xue
\cite {TX} where boson stars are studied in the framework of  a
scalar-tensor theory with a coupling between the dilaton and the
mass term of the boson field.

More recently boson stars in scalar-tensor theories have been
studied in the papers by Torres et al \cite{TLS}, \cite{TSL} and
in the paper by Comer and Shinkai \cite{CS}. In \cite{TLS} boson
stars have been studied in connection with so-called gravitational
memory \cite{Barrow}, while their stability through cosmic history
has examined in \cite{CS} and \cite{TSL}. The dynamical evolution
of boson stars has been investigated in the paper by Balakrishna
and Shinkai \cite{BS}. Charged boson stars in scalar-tensor
gravity with massless gravitational scalar have been studied in
\cite{WT}.

As we already mentioned boson stars in scalar-tensor theories have
been studied only in the case of scalar-tensor theories of gravity
with massless gravitational scalar. From a field-theoretical point
of view it is to be expected that the scalar part of the
gravitational field is massive and leads to a force with a finite
range. It also should be noted that there is no symmetry which
forbids a mass term, or a more general potential term for the
gravitational scalar. Moreover, if string theory and it's low
energy limit are relevant to the real world then the dilaton must
be massive. Unfortunately, our current understanding of how the
dilaton acquires  mass is rather primitive and it's tied to our
lack of understanding of supersymmetry breaking. Since we don't
have a definite model for dilaton  mass generation the mass of the
dilaton and the form of its potential are completely unknown from
theoretical point of view.

The purpose of the present paper is to study  boson stars in the
framework of some class of scalar-tensor theories of gravity with
massive gravitational scalar. In particular we study the possible
influence of the mass and more generally of the potential of
gravitational scalar on the equilibrium configurations and
stability of the boson stars.  The obtained results, however, are
typical  for wider classes of scalar-tensor theories with massive
dilaton not only for the class we consider.

\section{A scalar-tensor theory of gravity with massive gravitational
scalar}

The most general scalar-tensor theory of gravity including such a term
is described by the following action
\ben
\la{GAC}
{\tilde {\cal A}} = -{1 \over 16\pi G_{*}}\!
\int d^4x \sqrt{-{\tilde g}}
\left(\Phi {\tilde R} -
h(\Phi){\tilde g}^{\mu\nu}\partial_{\mu}\Phi \partial_{\nu}\Phi  \right.\nonumber \\
\left. +{\tilde U}(\Phi)\right)
 + {\cal A}_{matter}(\Psi_{matter}, {\tilde g}_{\mu\nu})
\een where ${\tilde g}_{\mu\nu}$  is the space-time metric,
${\tilde R}$ is the Ricci scalar curvature with respect to the
space-time metric. The functions $h(\Phi)$ and ${\tilde U}(\Phi)$
are, in general, arbitrary differentiable  functions of the
gravitational scalar $\Phi$. A new method for determination of
these functions
 using astrophysical observations is developed in the recent articles \cite{Starobinsky},
but at present we have no good enough data to make use of this method.
The last term in (\ref{GAC}) denotes the action of the matter, which is a
functional of the matter variables, collectively denoted by $\Psi_{matter}$,
and of the space-time metric ${\tilde g}_{\mu\nu}$.

For our purpose in this article the action (\ref{GAC}) is too
general. That's why we restrict ourselves to a subclass of
scalar-tensor theories which don't include explicitly a kinetic
term for the gravitational scalar in the action (i.e. with
function $h(\Phi) \equiv 0$). In particular we consider
scalar-tensor theories described by the action \ben \la{CAC}
{\tilde A} = {-1 \over 16\pi G_{*}}\! \int d^4x \sqrt{-{\tilde g}}
\left(\Phi {\tilde R} + {\tilde U}(\Phi)\right)    \nonumber\\ +
{\cal A}_{matter}(\Psi_{matter}, {\tilde g}_{\mu\nu}). \een Such
scalar-tensor theory with only one unknown function ${\tilde
U}(\Phi)$ is much more definite. It has been considered at first
by O'Hanlon \cite{OH} in connection with Fujii's theory of massive
dilaton \cite{Fujii1}. It may be considered also as a part of more
general theories of gravity like (4+1)-dimensional Kaluza-Klein
model described by Fujii in \cite{Fujii2}, or model of gravity
with violated local conformal symmetry in affine-connected spaces
which probably may be related with string theory \cite{Fiziev1}. A
recent development of  model of dilatonic gravity with action
(\ref{CAC}) related with newest astrophysical data on cosmological
constant problem may be found in \cite{Fiziev2}. In addition there
it was shown that this model is consistent with extremely high
precision with all known data in solar system gravitational
experiments if one chooses properly the potential ${\tilde
U}(\Phi)$. Under proper normalization of this potential solar
system end Earth surface experiments actually give restrictions
only on the mass term in the potential ${\tilde U}(\Phi)$.

The corresponding field equations are
\begin{eqnarray}
{\tilde G}_{\mu\nu} \! = \! {\kappa_{*}\over \Phi} {\tilde
T}_{\mu\nu} \! + \! {1\over \Phi}\! \left(\! {\tilde \nabla}_{\mu}
{\tilde \nabla}_{\nu}\Phi\! -\! {\tilde g}_{\mu\nu} {\tilde
\nabla}_{\rho} {\tilde \nabla}^{\rho}\Phi \! \right)\! +\! {1\over
2}\!\!\!\!&&\!\!\!\! {{\tilde U}(\Phi) \over \Phi}{\tilde
g}_{\mu\nu},\nonumber \\ [-0.5mm]\label{eq:1}\\[-0.5mm] {\tilde
\nabla}_{\rho}{\tilde \nabla}^{\rho}\Phi   + {1\over 3} \left(\Phi
{d{\tilde U}(\Phi) \over d\Phi} - 2{\tilde U}(\Phi) \right)\!\! &
= &\!\!{\kappa_{*}\over 3} {\tilde T}.\nonumber \een For the
treatment of the local Earth surface, astrophysical  and
astronomical problems in star systems (but not for cosmological
problems in the scales of universe)  we demand that a weak-field
approximation should be possible assuming in addition that the
space-time is asymptotically flat. Then in weak-field
approximation we may write

\ben {\tilde g}_{\mu\nu} = \eta_{\mu\nu} + \delta {\tilde
h}_{\mu\nu}, \,\,\,\,\, \,\,\, \Phi = \Phi_{\infty}  + \delta \Phi
\een where $\eta_{\mu\nu}$  is the flat space-time metric,
$\Phi_{\infty}$ is the background value of the $\Phi$, and $\delta
h_{\mu\nu}$ and $\delta \Phi$  are small perturbations due to the
local masses. Consistency of the field equations then requires
that \ben \la{CC} {\tilde U}(\Phi_{\infty}) = {d{\tilde U}\over
d\Phi}(\Phi_{\infty}) = 0. \een Using now the consistency
conditions (\ref{CC})  we obtain that the mass of the
gravitational scalar $\Phi$ is \ben m^{2} = {\Phi_{\infty} \over
3} {d^2 {\tilde U}\over d\Phi^2}(\Phi_{\infty}). \een Further on,
without loss of generality we take $\Phi_{\infty}=1.$ In this
case, taking into account that the range of the gravitational
scalar is finite we obtain that  the bare gravitational constant
$G_{*}$ coincides  with the background physical gravitational
constant.

The PPN formalism for scalar-tensor theories of gravity with a
massive gravitational scalar has been developed by Helbig
\cite{H}, who has also calculated the basic gravitational effects
as perihelion shift and so on. Using Helbig's results and the
experimental data published in \cite{ADR} and \cite{TA}  we obtain
that the scalar-tensor theory we consider here is consistent with
the experiments when the Compton length of the gravitational
scalar $\Phi$ is less than $2$ mm,  or equivalently when  the mass
of the gravitational scalar satisfies the constraint $m \ge
10^{-4} eV$ \cite{Fiziev2}.

So far our considerations have been made in the Jordan frame. For
some purposes, in particular for numerical calculations, the
scalar-tensor theories are better formulated in the conformal
Einstein frame. The metric in the Einstein frame is given by the
conformal transformation \ben g_{\mu\nu} = \Phi {\tilde
g}_{\mu\nu} \; . \een It should be stressed that it's the  Jordan
frame which is the physical frame while the Einstein frame is a
convenient mathematical tool. Including also the dilaton field
$\varphi$ via the formula $\Phi = \exp(-2\alpha \varphi)$, where
$\alpha = {1\over \sqrt{3}}$, the action (\ref{CAC}) written in
terms of $g_{\mu\nu}$ and $\varphi$ reads \ben {\cal A} = -{1
\over 16\pi G_{*}} \int d^4x \sqrt{-g} \left(R -
2g^{\mu\nu}\partial_{\mu}\varphi\partial_{\nu}\varphi +
U(\varphi)\right)  \nonumber\\ + {\cal A}_{matter}(\Psi_{matter},
A^2(\varphi)g_{\mu\nu}). \een Here $R$ is the Ricci scalar
curvature with respect to the metric $g_{\mu\nu}$, $U(\varphi) =
A^4(\varphi){\tilde U}(\Phi(\varphi))$ is the potential of the
dilaton field $\varphi$ and $A^2(\varphi) = \Phi^{-1}(\varphi)$.
It should be noted that the convention $\Phi_{\infty} = 1$ in
terms of the dilaton field writes $\varphi_{\infty} = 0$. As must
be expected the dilaton potential $U(\varphi)$  satisfies the
conditions $U(\varphi_{\infty}) = {dU(\varphi) \over
d\varphi}(\varphi_{\infty}) = 0$  and the mass of the dilaton
coincides with the mass of the gravitational scalar $\Phi$ which
gives another equivalent representation of the same physical
object.

\section{Boson star model}

We take the matter action to be the action of complex, massive and
self-interacting scalar field $\Psi$ which in the physical Jordan
frame has the form \ben {\cal A}_{matter}\! =\! \int \!\!d^4x
\sqrt{-{\tilde g}} \left({1\over 2} {\tilde
g}^{\mu\nu}\partial_{\mu}\Psi^{+}\partial_{\nu}\Psi -
W(\Psi^{+}\Psi)\right) \een where $W(\Psi^{+}\Psi)\! =\! {1\over
2}m^2_{B}\Psi^{+}\Psi\!  +\! {1\over 4}{ \hat
\Lambda}(\Psi^{+}\Psi)^2$  is the potential of the boson field.
Then the action of the gravity and matter in Einstein frame is
\begin{eqnarray}
{\cal A} = -{1\over 16\pi G_{*}} \int \!&&\!d^4x \sqrt{-g}
\left(R - 2g^{\mu\nu}\partial_{\mu}\varphi \partial_{\nu}\varphi
+ U(\varphi) \right) \nonumber\\
+\!\int\! d^4x \sqrt{-g}&& \left({1\over 2}A^2(\varphi)g^{\mu\nu}\partial_{\mu}\Psi^{+}
\partial_{\nu}\Psi \right.\nonumber \\
 \bigg.&&\hspace{1cm}- A^4(\varphi)W(\Psi^{+}\Psi) \bigg) .
\la{EFBSA}
\end{eqnarray}
Varying the action (\ref{EFBSA})  with respect to $g_{\mu\nu}$,
$\varphi$, $\Psi$  and $\Psi^{+}$  we obtain the field equations
\ben \la{EFFE} G_{\mu\nu} = \kappa_{*}T_{\mu\nu}  +
2\partial_{\mu}\varphi\partial_{\nu}\varphi  -
g_{\mu\nu}\partial_{\rho}\varphi\partial^{\rho}\varphi  + {1\over
2}U(\varphi)g_{\mu\nu},  \\   \nonumber
\nabla_{\rho}\nabla^{\rho}\varphi  + {1\over 4}
U^{\prime}(\varphi) = -{\alpha \over 2}\kappa_{*} T,  \\ \nonumber
\nabla_{\rho}\nabla^{\rho}\Psi + 2\alpha
\partial_{\rho}\varphi\partial^{\rho}\Psi = -2A^2(\varphi)
{dW(\Psi^{+}\Psi) \over d\Psi^{+}}, \\ \nonumber
\nabla_{\rho}\nabla^{\rho}\Psi^{+} + 2\alpha
\partial_{\rho}\varphi\partial^{\rho}\Psi^{+} = -2A^2(\varphi)
{dW(\Psi^{+}\Psi) \over d\Psi} \een where $\nabla_{\rho}$ is the
Levi-Civita connection with respect to the metric $g_{\mu\nu}$,
$T_{\mu\nu}$  is the Einstein frame energy-momentum tensor  of the
boson filed $\Psi$  and $T$ is its trace. The Einstein frame
energy-momentum tensor $T_{\mu\nu}$ is given by \ben \la{EFEMT}
T_{\mu\nu} = {1 \over 2} A^2(\varphi)
\left(\partial_{\mu}\Psi^{+}\partial_{\nu}\Psi  +
\partial_{\mu}\Psi\partial_{\nu}\Psi^{+} \right)   -    \\  \nonumber
{1\over
2}A^2(\varphi)\left(\partial_{\rho}\Psi^{+}\partial^{\rho}\Psi -
2A^2(\varphi)W(\Psi^{+}\Psi)\right)g_{\mu\nu}. \een We consider a
static and spherically symmetric boson star. Then the metric
$g_{\mu\nu}$  can be written in the standard form \ben \la{EFMET}
ds^2 = e^{\nu(\rho)} dt^2  - e^{\lambda(\rho)} d\rho^2 - \rho^2
\left(d\theta^2 + \sin^2(\theta) d\phi^2 \right). \een We also
demand a spherically-symmetric form for the boson field and we
adopt a form  consistent with the static metric: \ben \la{SFBF}
\Psi ={\hat \sigma}(\rho) e^{i\omega t}. \een Using the metric
(\ref{EFMET}) and the equation defining the form of the boson
field  together   with the Einstein frame energy-momentum tensor
(\ref{EFEMT})  in the field equations (\ref{EFFE}) we get the
equations for the structure of the boson star. Before  we
explicitly write them we are going to introduce a dimensionless
radial coordinate by \ben r = m_{B}\rho. \een From now on, a prime
will denote a differentiation with respect to the dimensionless
coordinate $r$. We also define other dimensionless quantities by
\ben \Omega = {\omega \over m_{B}} , \; \sigma = \sqrt{\kappa_{*}}
{\hat \sigma}, \; \Lambda = {{\hat \Lambda} \over \kappa_{*}
m_{B}}, \; \gamma = {m\over m_{B}} \een and dimensionless
potential $V(\varphi)$ given by \ben U(\varphi) = m^2 V(\varphi).
\een With all these definitions, the equations of the structure of
the boson star in Einstein frame reduce to the following \ben
\la{EFODE} \lambda^\prime = {1 - e^{\lambda}\over r} +
re^{\lambda}{\cal T}^{0}_{0} + r {\varphi^\prime}^2 + {1\over
2}\gamma^2 re^{\lambda}V(\varphi), \\ \nonumber \nu^{{\,}\prime} =
{e^{\lambda} - 1 \over r} -  re^{\lambda}{\cal T}^{1}_{1} + r
{\varphi^\prime}^2 - {1\over 2}\gamma^2 re^{\lambda}V(\varphi),\\
\nonumber \nu^{{\,}\prime\prime}\!=
\!{-{\nu^{{\,}\prime}}^2\!+\!\lambda^{{\,}\prime}\!-\!\nu^{{\,}\prime}\!+\!\lambda^{{\,}\prime}\nu^{{\,}\prime}\over
2} \!-\!2e^{\lambda} {\cal
T}^{2}_{2}\!-\!2{\varphi^{\prime}}^2\!-\! \gamma^2
e^{\lambda}V(\varphi), \\ \nonumber \varphi^{\prime\prime} =
-\left({\nu^{{\,}\prime} - \lambda^\prime \over 2}  + {2\over r}
\right) \varphi^\prime  + {1\over 4}\gamma^2 e^{\lambda}
{dV(\varphi)\over d\varphi}  + {\alpha \over 2}{\cal
T}e^{\lambda}, \\ \nonumber \sigma^{\prime\prime} = -
\left({\nu^{{\,}\prime} - \lambda^\prime \over 2}  + {2\over r}
\right) \sigma^\prime  -  \Omega^2 e^{\lambda - \nu} \sigma -
2\alpha \varphi^\prime \sigma^\prime  +    \\ \nonumber
2A^2(\varphi)e^{\lambda}{dW(\sigma^2)\over d\sigma^2} \sigma. \een
Here ${\cal T}^{\mu}_{\nu} = {\kappa_{*}\over m^2_{B}}
T^{\mu}_{\nu}$ is the  dimensionless Einstein frame
energy-momentum tensor  and ${\cal T}$ is its trace. In explicit
form we have \ben {\cal T}^{0}_{0}\!=\!{1\over 2}\Omega^2
A^2(\varphi)e^{-\nu}\sigma^2 \!+\!{1\over
2}A^2(\varphi)e^{-\lambda} {\sigma^\prime}^2\!+\!
A^4(\varphi)W(\sigma^2),\\ \nonumber {\cal
T}^{1}_{1}\!=\!-\!{1\over 2}\Omega^2
A^2(\varphi)e^{\!-\!\nu}\sigma^2 \!-\!{1\over
2}A^2(\varphi)e^{\!-\!\lambda} {\sigma^\prime}^2\!+\!
A^4(\varphi)W(\sigma^2), \\  \nonumber {\cal
T}^{2}_{2}\!=\!-\!{1\over 2} \Omega^2
A^2(\varphi)e^{\!-\!\nu}\sigma^2 \!+\!{1\over
2}A^2(\varphi)e^{\!-\!\lambda} {\sigma^\prime}^2\!+\!
A^4(\varphi)W(\sigma^2) , \\  \nonumber {\cal T}\!=\!-\!\Omega^2
A^2(\varphi)e^{-\nu}\sigma^2   + A^2(\varphi)e^{-\lambda}
{\sigma^\prime}^2  + 4 A^4(\varphi)W(\sigma^2).\!\!\!\!\! \een By
reasons imposed by the numerical method used in the present paper,
instead to investigate the system (\ref{EFODE}) we solve
numerically an equivalent system obtained as follows. First making
use of the second equation of (\ref{EFODE}) we express
$e^{\lambda}$ as a function of $\nu, \nu^{{\,}\prime}, \sigma,
\sigma^{\prime}, \varphi, \varphi^{\prime}$ and then substitute in
the other equations. In this way we obtain the system: \ben
\la{EFODE1} \nonumber \nu^{{\,}\prime\prime} = - {\nu^{{\,}\prime}
\over r}   + \left(- {\nu^{{\,}\prime} \over r} + {\cal T}^{0}_{0}
- {\cal T}^{1}_{1} + 2{\cal T}^{2}_{2}   - \gamma^2 V(\varphi)  +
\right. \\ \nonumber \left. {\nu^{{\,}\prime} r \over 2}
\left({\cal T}^{0}_{0} + {\cal T}^{1}_{1}  + \gamma^2
V(\varphi)\right) \right)e^{\lambda}, \een \ben
\varphi^{\prime\prime} = - {\varphi^{\prime} \over r}   + \left(-
{\varphi^{\prime} \over r} + {\alpha \over 2}{\cal T} + {1\over 4}
\gamma^2 { dV(\varphi)\over d\varphi }  +  \right.\\ \nonumber
\left. {\varphi^{\prime} r \over 2} \left({\cal T}^{0}_{0} + {\cal
T}^{1}_{1}  + \gamma^2 V(\varphi)\right) \right)e^{\lambda}, \een
\ben \nonumber \sigma^{\prime\prime} = - {\sigma^{\prime} \over r}
- 2\alpha \varphi^{\prime} \sigma^{\prime}   + \left( -
{\sigma^{\prime} \over r}  + 2A^2(\varphi){dW(\sigma^2)\over
d\sigma^2} \sigma -  \right. \\ \nonumber \left. \Omega^2 e^{-\nu}
\sigma +  {\sigma^{\prime} r \over 2} \left({\cal T}^{0}_{0} +
{\cal T}^{1}_{1}  + \gamma^2 V(\varphi)\right) \right)e^{\lambda}
\een where $e^{\lambda}$  is given by \ben e^{\lambda}\!=\!\!{1+
r\nu^{{\,}\prime} - r^2{\varphi^{\prime}}^2 - {1\over 2}
A^2(\varphi)r^2 {\sigma^{\prime}}^2 \over
1\!-\!r^2\!\left({\!1\over 2}\gamma^2 V\!(\varphi)\!-\! {1\over
2}\Omega^2
A^2(\varphi)e^{\!-\!\nu}\sigma^2\!+\!A^4(\varphi)W\!(\sigma^2)
\right)}.\!\! \een We solve numerically the system (\ref{EFODE1}).
The boundary conditions for the system are the following. We
demand asymptotic flatness which means that $\nu(\infty)=0$. The
non-singularity of the $\nu$ at the center of the star requires
that $\nu^{{\,}\prime}(0) = 0$. Concerning the dilaton $\varphi$,
the non-singularity at the center implies $\varphi^{\prime}(0)=0$,
while  at infinity we must have $\varphi(\infty) = 0$ as required
by the asymptotic flatness. Non-singularity of the boson field at
the center implies $\sigma^{\prime}(0) = 0$ and the asymptotic
flatness requires that $\sigma(\infty) = 0$. To complete the
nonlinear eigenvalue problem for $\Omega$ we also must give the
central value of the boson field $\sigma(0) = \sigma_{c} $.

Here we must make some comments. We have imposed the boundary
conditions in Einstein frame. But what is important is the
non-singularity and asymptotic flatness in the physical Jordan
frame. However, it's not difficult to see that all Einstein frame
boundary conditions phrased in Jordan frame remain the same.
Indeed, the only quantity which changes under transition to Jordan
frame is ${\nu}$. The non-singularity at the center and the
asymptotic flatness in Jordan frame require correspondingly that
${\tilde \nu}^{\prime}(0)= 0$ and ${\tilde \nu}(\infty)=0$. Taking
into account that ${\tilde \nu}= \nu + 2\alpha \varphi$ and
Einstein frame boundary conditions for the dilaton field $\varphi$
we obtain that Jordan frame boundary conditions for ${\tilde \nu}$
are satisfied if and only if ${\nu}^{\prime}(0)= \nu (\infty)= 0$.

\section{Conserved quantities}
The action of the boson field is $U(1)$-invariant under the global
gauge transformations $\Psi \to e^{ia}\Psi$, where $a$  is a
constant. This global $U(1)$-symmetry gives rise to the following
conserved current in Jordan frame

\ben
{\tilde J}^{\mu} = {i\over 2}\sqrt{-{\tilde g}} {\tilde
g}^{\mu\nu} \left(\Psi\partial_{\nu}\Psi^{+}  -
\Psi^{+}\partial_{\nu}\Psi \right).
\een

The same current written in terms of the Einstein frame metric  is

\ben
J^{\mu} = {i\over 2}A^2(\varphi)\sqrt{- g} g^{\mu\nu}
\left(\Psi\partial_{\nu}\Psi^{+}  - \Psi^{+}\partial_{\nu}\Psi
\right).
\een

The conserved current leads to conserved charge - the total number
of particles making up the star:

\ben
N = \int {\tilde J}^{0} d^3x.
\een

It should be stressed that $N$ is the total number of particles in
the physical Jordan frame. Taking into account the explicit form
of the metric and of the boson field we obtain

\ben N = \omega \int d^3x  \sqrt{- {\tilde g}}\, {\tilde g}^{00}\,
 {\hat\sigma}^2 \een

In order to calculate numerically the total  number of particles
we must present the above integral in terms of Einstein frame
metric. This can be done easily by presenting the Jordan frame
metric as conformally transformed Einstein frame metric. This way
we obtain

\ben {\cal M}_{R} = m_{B} N = \left({ {{\cal M}^2}_{Pl} \over
2m_{B}} \right) \Omega \int_{0}^{\infty} dr r^2 A^2(\varphi)
e^{{\lambda - \nu \over 2}} \sigma^2 \een where ${\cal M}_{R}$  is
the rest mass of the star in the Jordan frame.

The binding energy in Jordan frame is then defined by \ben {\cal
E}_{b} = {\cal M} - {\cal M}_{R} \een where ${\cal M}$ is the
total mass of the star.

In contrast to the general relativity the definition of mass in
scalar-tensor theories is subtle . There are three possible mass
definitions in Jordan frame, notably the Schwarzschild mass ${\cal
M}_{S}$ (i.e. ADM mass in Jordan frame), the Kepler  mass ${\cal
M}_{K}$ and the tensor mass ${\cal M}_{T}$ which is the ADM mass
in Einstein frame \cite{W}. These mass definitions are equivalent
in general relativity, but they may give different results in
scalar-tensor theories of general type. It's the tensor mass which
has got physically acceptable properties: The tensor mass has the
important property to peak, as a function of the central density
at the same location where the particle number (respectively the
rest mass) takes its maximum. This property has been proved in the
most general scalar-tensor theory including a potential term for
the gravitational scalar (dilaton), see for details \cite{Y}. That
the tensor mass and particle number peak at the same location
results in a cusp in the bifurcation diagram ${\cal M}$
(respectively ${\cal E}_{b}$) versus ${\cal M}_{R}$.

All masses we have mentioned are asymptotically measured
quantities. Taking into account that in the scalar-tensor theory
we consider the gravitational scalar (dilaton) varies essentially
in a finite space-time domain , it's not difficult to see that all
masses in our case are identical. That's why all properties proven
for the tensor mass are shared also by the others and the mass of
the star may be expressed in three different ways as one wishes.
We take the expression for the mass of the star in the form of the
Schwarzschild mass  in Einstein frame, namely \ben {\cal M} =
\left({ {{\cal M}^2}_{Pl} \over 2m_{B}} \right) \int_{0}^{\infty}
dr r^2 \left({\cal T}^{0}_{0}  + e^{-\lambda} {\varphi^\prime}^2 +
{1\over 2}U(\varphi) \right). \een

For numerical purposes we also introduce the following
dimensionless masses \ben M = {{\cal M} /\,\,\, {{\cal M}^2_{Pl}
\over 2m_{B}}} \,\,\,\,\,\,\,\,\,  ,   \,\,\,\,\,\,\, M_{R} =
{{\cal M}_{R} /\,\,\, {{\cal M}^2_{Pl}\over 2m_{B}}} \, \,\, . \een
Therefore the dimensionless binding energy is \ben E_{b}= M -
M_{R}\, \, . \een

We also define an effective dimensionless radius of the boson star as:
\ben
R = {m_{B}\over N} \int r J^{0} d^3x \, .
\een

\section{Numerical results and discussions}
To solve the nonlinear ODE's (\ref{EFODE1})  with the
corresponding boundary conditions  and spectral parameter $\Omega
$ we apply the continuous analog of the Newton method
\cite{gavurin}, \cite{jidkov}. In our case this method leads to a
linear systems ODE's with respect to the increments of unknown
functions coupling with an algebraic equation for the increment of
$\Omega $. The appropriate linear boundary value problems we solve
numerically using the spline collocation at the Gaussian points on
the irregular mesh condensing to the centre $r=0$ of the star. For
details see our  work \cite{BTFY}.

As we have already mentioned at present we have only some weak
experimental constrains on the mass of the dilaton and on the form
of its potential. A definite theoretical suggestions for them do
not exist. That's why we consider  first the simplest form for the
potential of the gravitational scalar \ben {\tilde U}(\Phi) =
{3\over 2}m^2 \left(\Phi - 1 \right)^2. \een The corresponding
dilaton potential is \ben U(\varphi) = m^2 V(\varphi) = {3 \over
2}m^2 \left(1 - \exp({2\varphi \over \sqrt{3}}) \right)^2. \een
The numerical results for the chosen potential are presented in
the figures.

\begin{figure}%1
{\epsfig{file=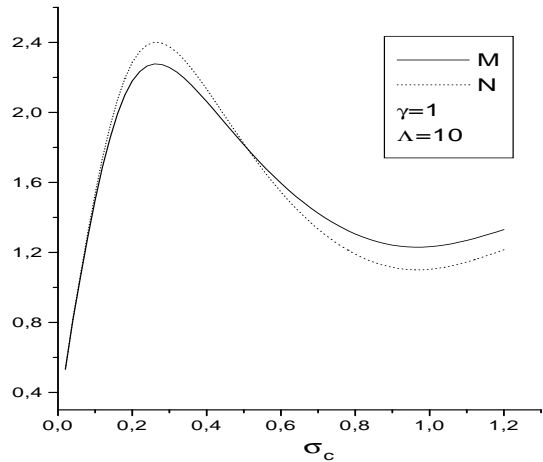, , width=3.1in, height=3.1in, scale=1.31, bb=
10 10 550 850}} \caption{ The dependence of the mass and rest mass
of the boson star on the central value $\sigma_{c}$ }
\end{figure}

Fig.1 presents a configuration diagram which is typical for large
variations of the parameters not only for those values presented
on the figure.

\begin{figure}%2
{\epsfig{file=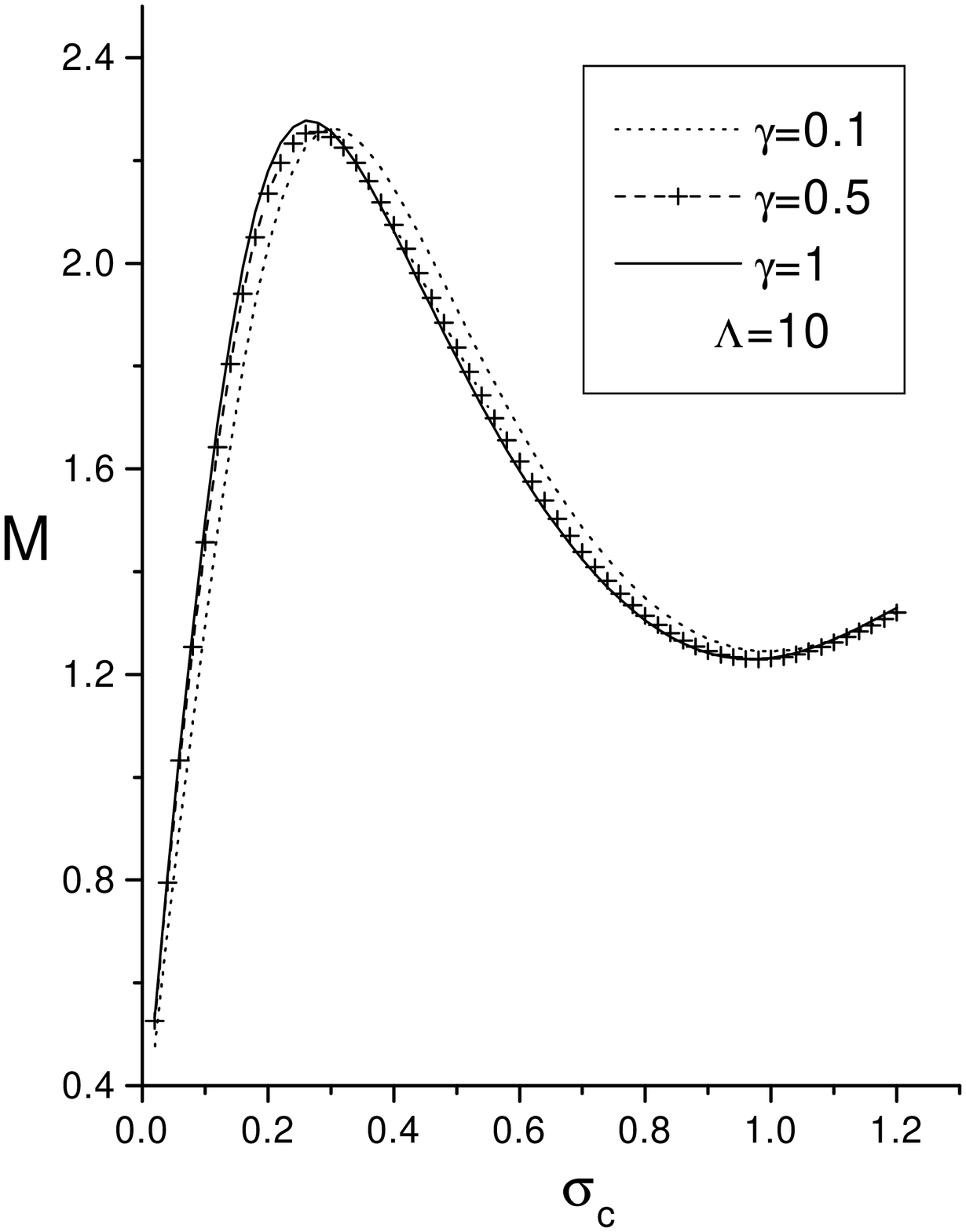, , width=3.1in, height=3.1in, scale=1.31,
bb=10 10 550 850}} \caption{The dependence of the boson star mass
on the central value $\sigma_{c}$ for three different values of
the parameter $\gamma$.}
\end{figure}
The dependence of the boson star mass on the central value
$\sigma_{c}$ is presented in Fig.2 for three different values of
the ratio $\gamma= {m \over m_{B}}$ at fixed value of the
parameter $\lambda = 10.$ It's seen that the curves
$M(\sigma_{c})$ differ slightly from each other. More detailed
investigation for much more values of $\gamma$ shows that the
curves tend to the same curve $M(\sigma_{c})$ from the general
relativistic case for large values of $\gamma$ ($\gamma \approx
10$). This behaviour is confirmed also by Fig.3 where the
dependence of the boson star mass on the parameter $\gamma$ is
shown for $\sigma_{c} = 0.1$ and $\Lambda = 10$.
\begin{figure}[h]%3
{\epsfig{file=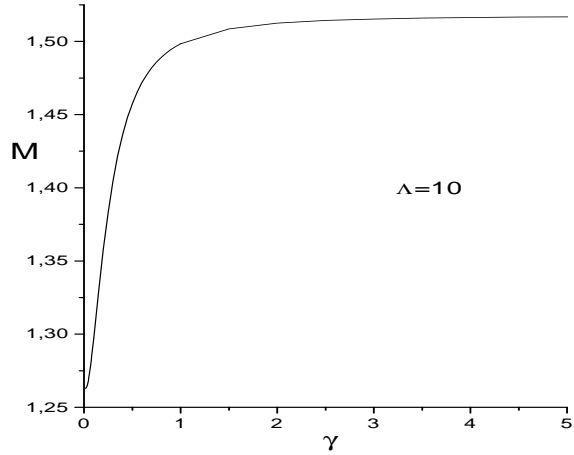, width=3.1in, height=3.1in, scale=1.31, bb=10
10 550 850}} \caption{ The dependence of the boson star mass on
the parameter $\gamma$ for $\Lambda = 10$ }
\end{figure}
For small values of $\gamma$ the boson star mass increases when
the parameter $\gamma$ increases. Then for larger values of
$\gamma$ the mass $M$ tends to a fixed value which is just the
general relativistic value.  The cause for this behaviour is that
the increasing of the parameter $\gamma$ is actually an increasing
of the dilaton mass with respect to the boson mass which leads to
smaller dilaton range and  therefore to a suppressing of  the
dilaton field which  means that the general relativity is
recovered. The Fig. 3 also shows that the boson stars in the model
under consideration always have masses smaller than those in
general relativity.  The deviations from general relativity are
within a typical few percent.

\begin{figure}%4
{\epsfig{file=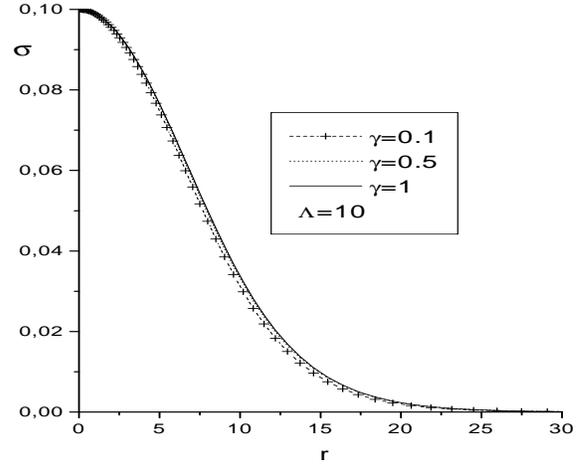, , width=3.1in, height=3.1in, scale=1.31,
bb=10 10 550 850}} \caption{ The dependence $\sigma(r)$ for three
different values of the parameter $\gamma$.  }
\end{figure}

 The influence of the dilaton mass, respectively $\gamma$, on the
dependence of the boson field on the radial coordinate $r$ is very
slightly as one may see from Fig. 4. The dependence $\sigma(r)$ is
presented there for three  different values of $\gamma$. The three
curves are extremely close to each other.

The dependence ${\tilde \nu}(r)$ is shown  in Fig. 5 for three
different values of $\gamma$. As it is seen  the influence of
$\gamma$ on ${\tilde \nu(r)}$ is slight.

\begin{figure}[h]%5
{\epsfig{file=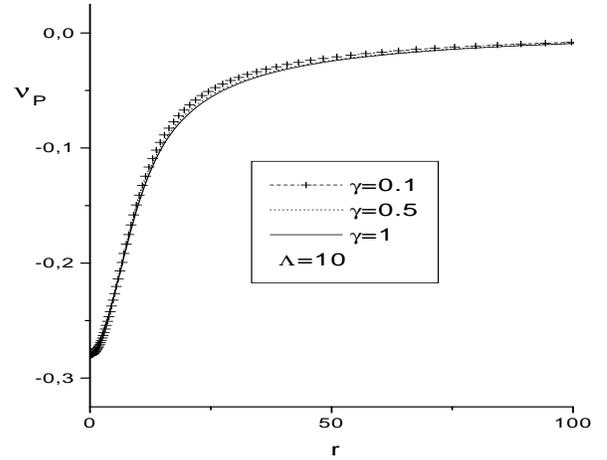, , width=3.1in, height=3.1in, scale=1.31,
bb=10 10 550 850}} \caption{  The physical function ${\tilde
\nu(r)}$   for three different values of the parameter $\gamma$. }
 \end{figure}
\begin{figure}[h]%6
{\epsfig{file=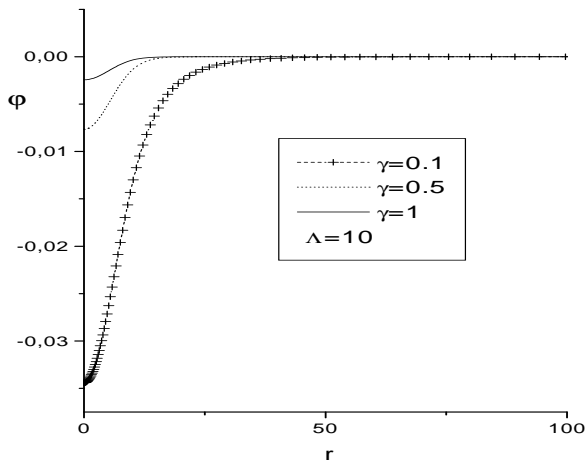, , width=3.1in, height=3.1in, scale=1.31,
bb=10 10 550 850}} \caption{The dependence $\varphi(r)$ for three
different values of the parameter $\gamma$.  }
\end{figure}
The most sensitive to the value of  the parameter $\gamma$ is the
dilaton field $\varphi$. The dependence $\varphi(r)$ is presented
in Fig. 6 for three different values of $\gamma$. One sees that
the increasing of $\gamma$ leads to suppressing the dilaton field
as one must expect. The dependence $\Phi(r)$ is shown in Fig. 7.
At the center of the star one has $\Phi(0) > 1$ which means that
the gravitational constant at the center satisfies $G(0) < 1 $.
Therefore the boson stars are less gravitationally bounded than in
general relativistic case.

\begin{figure}%7
{\epsfig{file=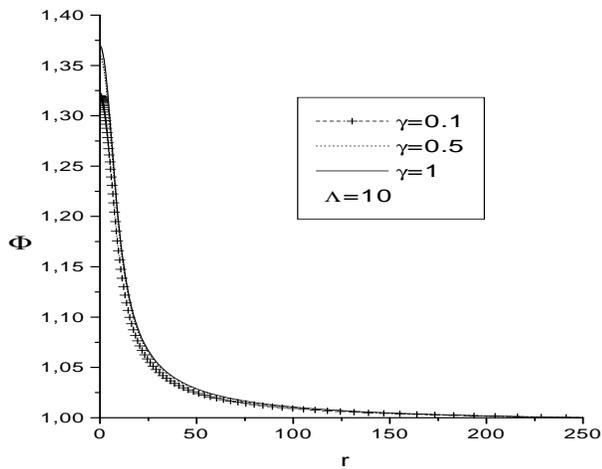, , width=3.1in, height=3.1in, scale=1.31,
bb=10 10 550 850}} \caption{ The function $\Phi(r)$  for three
different values of the parameter $\gamma$}
\end{figure}

We have also examined the cases of different values of $\Lambda$.
The picture is qualitatively the same.

It is interesting to know how the exact form of the dilaton
potential $V(\varphi)$ influences the boson star structure. We
have examined several  potentials $V(\varphi)$  with the same
dilaton mass (for example $V(\varphi)= 2\varphi^2$, $V(\varphi)= 2\sin^2(\varphi)$,
$V(\varphi)= 4(1- \cos(\varphi))$, $V(\varphi) = 2(1 - \exp(-\varphi^2))$). Our numerical results show that boson star structure
in practice does not depend on the exact form of the dilaton
potential in the class of potential we have examined. This gives
strong evidences that the boson star structure is sensitive mainly
to the second derivative of the dilaton potential ${d^2
V(\varphi)\over d\varphi^2}(0)$ determining the dilaton mass
rather to the exact form of the potential.

\begin{figure}[h]%8
{\epsfig{file=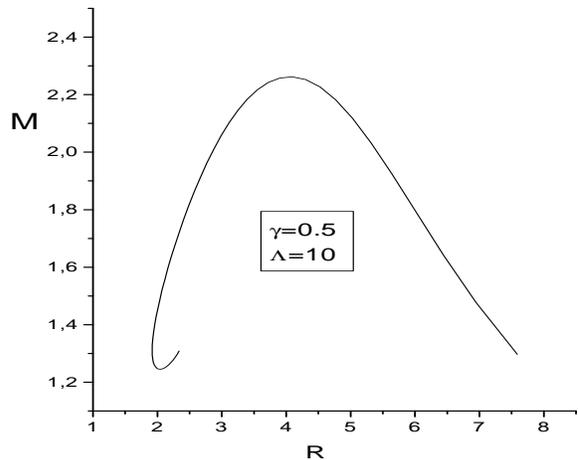, , width=3.1in, height=3.1in, scale=1.31,
bb=10 10 550 850}} \caption{ The boson star mass $M$ as a function
of the boson star radius $R$}
\end{figure}
Finally in Fig. 8 one can see that the usual dependence of the
mass $M(R)$ of boson star on its effective radius $R$  take place
in the scalar-tensor model with massive dilaton, too. As seen,
there exist a domain of stability of the boson star, followed by
domains of unstable states.

\begin{figure}%9
{\epsfig{file=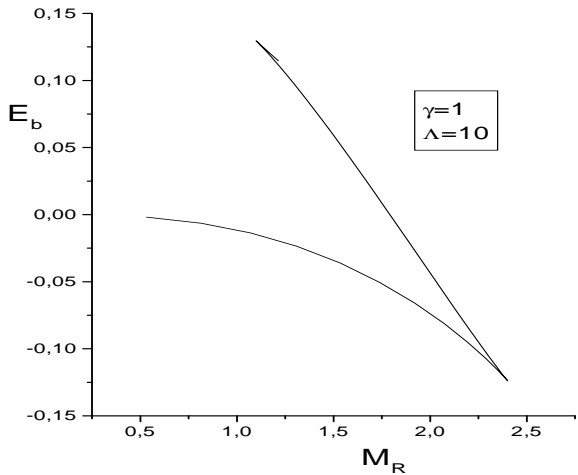, , width=3.1in, height=3.1in, scale=1.31,
bb=10 10 550 850}} \caption{ The boson star binding energy $E_{b}$
as a function of the rest mass $M_{R}$}
\end{figure}

\section{Boson star stability}

Here we briefly discuss stability property of the boson stars we consider.
Our analysis is based on catastrophe theory \cite{KMS}.

The basis for our analysis is Fig. 9 where the binding energy
$E_{b}$ is plotted against the rest mass $M_{R}$ (i.e. particle
number). This figure, as our numerical results show,  is typical
for a large  number of values of  $\gamma$ and $\Lambda$.
Therefore  the forthcoming conclusions  concern the general case
not  only   the special  case presented  by the figure.

Fig. 9 is actually a bifurcation diagram. For small central
densities the binding energy is negative which shows that the
boson star is potentially stable and one assumes that the boson
star is stable against small radial perturbations. As the central
density is increased one meets a cusp. The second branch as a
whole has higher mass and therefore it is unstable. At the cusp,
the boson star stability changes - one radial perturbation mode
develops instability.

\section{Conclusion}

In this paper we have analyzed static spherically symmetric boson
stars in a scalar tensor theory of gravity with massive
gravitational scalar (dilaton). We have studied their equilibrium
properties for different values of the ratio $\gamma= {m\over
m_{B}}$. The conclusion is that stable boson stars may exist for
any value of  $\gamma$. When $\gamma$ increases the equilibrium
configurations tends to those from general relativity and for
sufficiently large $\gamma$ the general relativistic case is in
practice recovered. The masses of the boson stars are turned out
to be always smaller than those  in general relativity. The
typical deviations of the masses in our model of dilatonic gravity
from those in general relativity is a few percent. The small
deviations of bosonic star structure in the considered  model of
dilatonic gravity from general relativistic case are significantly
greater then the corresponding deviations in solar system and
Earth surface experiments studied in \cite{Fiziev2}. It turns out
also that the boson star structure is sensitive mainly to the mass
term of the dilaton potential rather to the exact form of the
potential. The boson stars in our case are less gravitationally
bound as a consequence of the fact that the gravitational constant
$G$ within the stars is smaller than one.

For completeness we have examined also the case of scalar-tensor
theories with a massive gravitational scalar with kinetic term
$\omega_{BD}\,\Phi ^{-2}\,{\tilde g}^{\mu\nu}\partial_{\mu}\Phi
\partial_{\nu}\Phi$. In these cases the obtained results for
different values of the parameter $\omega_{BD}$ are qualitatively
the same as in the case $\omega_{BD}\equiv 0$ and for large
$\omega_{BD}$ the differences between bosonic stars in dilatonic
gravity and in general relativity are even smaller.

It is worth noting that, in the domain of stability of boson
stars, our results are qualitatively close to the results obtained
for boson stars in Brans-Dicke theory \cite{GJ}, \cite{T}, and for
boson stars considered in a scalar-tensor theory with a kinetic
term and  simple coupling between the dilaton and the mass term of
the boson filed in Jordan frame \cite{TX} which in our notations
reads $m^2_{B}\Phi\, \Psi^{+}\Psi$. In the domain of instability
of boson stars the results of the last reference differ
essentially from ours. As in general relativity and Brans-Dicke
theory beyond the point of stability for large central densities
the boson star mass in our model of massive dilatonic gravity
drops, oscillates a bit and approaches a constant value while in
the model considered in \cite{TX} it increases rapidly.

\bigskip
{\em Acknowledgments:} We are grateful to the unknown referee for
useful suggestions and for pointing out the references \cite{TX}
and \cite{WT}. This work was supported by Bulgarian National
Scientific Fund, Contr. NoNo F610/99, MM602/96 and by Sofia
University Research Fund, Contr. No. 245/99.


\begin{thebibliography}{}

\bibitem{K} D.~Kaup,
            Phys. Rev. {\bf 172}, 1331 (1968)
%
\bibitem{RB} R.~Ruffini, S.~Bonazzola,
             Phys. Rev. {\bf 187}, 1767 (1969)
%
\bibitem{CSW} M.~Colpi, S.~Shapiro, I.~Wasserman,
              Phys. Rev. Lett. {\bf 57}, 2485 (1986)
%
\bibitem{GJ}   M.~Gunderson, L.~Jensen,
               Phys. Rev. D{\bf D48}, 5628 (1993)
%
\bibitem{T}   D.~Torres,
                Phys. Rev. D{\bf D56}, 3478 (1997)

%
\bibitem {TX}  Z.~Tao, X.~Xue, Phys. Rev. {\bf D45}, 1878 (1992)
%
\bibitem{TLS}   Torres D, Liddle A, Schunck F,
                Phys. Rev. D{\bf 57}, 4821 (1998)
%
\bibitem{CS}   G. Comer, H. Shinkai,
                Class. Quantum Grav. {\bf 15}, 669 (1998)
%
\bibitem{TSL}   D. Torres, F. Schunck, A. Liddle,
                Class. Quantum Grav. {\bf 15}, 3701 (1998)
%
\bibitem{Barrow} J. Barrow,
                 Phys. Rev. D{\bf 46}, 3227 (1992)
%
\bibitem{BS}  J. Balakrishna,H. Shinkai,
                Phys. Rev. D{\bf 58}, 044016-1 (1998)

%
\bibitem{WT} A.~Whinnett, D.~Torres,
             Phys. Rev. {\bf D60}, 104050, (1999)
%
\bibitem{Starobinsky} A.~A.~Starobincky, JETP Lett. {\bf 68}, 757 (1998);
                                   astro-ph/9811360;
                                  V.~Sahni, A.~A.~Starobinsky, astro-ph/9904398;
                                   T.~D.~Saini, S.~Raychaudhury, V.~Sahni, A.~A.~Starobinsky, astro-ph/9910231;
                                   B.~Boisseau, G.~Esposito-Far\'ese, D.~Polarski, A.~A.~Starobinsky, gr-qc/0001066.
%
\bibitem{OH} J.~O'Hanlon, Phys. Rev. Lett. {\bf 29}, 137 (1972).
%
\bibitem{Fujii1} Y.~Fujii, Nature (London) Phys. Sci, {\bf 234}, 5 (1971);
                Ann. Phys. (NY) {\bf 69}, 494 (1972).
                J.~O'Hanlon, Phys. Rev. Lett. {\bf 29}, 137 (1972).
                Y. Fujii, Phys. Rev. D {\bf 9}, 874 (1974).
%
\bibitem{Fujii2}    Y. Fujii, {\em Gravitation and scalar field}, Kodan-sha,
                   Tokyo, 1997.
%
\bibitem{Fiziev1}    P.~Fiziev, {\em Torsion Dilaton and Novel Minimal Coupling
                  Principle}, E-print: gr-qc/9809001;
%
\bibitem{Fiziev2}    P.~Fiziev, {\em A Realistic Model of Dilatonic Gravity}, E-print: gr-qc/9911037.
%
\bibitem{H} T.~Helbig, ApJ {\bf 382}, 223 (1991).
%
\bibitem{ADR} A.~De~R\'ujula, Phys. Lett. B {\bf  180}, 213 (1986).
%
\bibitem{TA} C.~Talmage, {\em in Proc. Sixth Marcel Grossman Meeting on General Relativity,}
ed. H.~Sato and T.~Nakamura, World Scientific, Singapore (1992)
%
\bibitem{W} A.~Whinnett, Class. Quant. Grav. {\bf 16}, 2796  (1999)
%
\bibitem{Y} S. ~Yazadjiev ,
            Class. Quantum Grav. {\bf 16}, L63 - L69  (1999)
%
\bibitem{gavurin}  M.~ Gavurin,  Izvestia VUZ, Matematika,{\bf  No 5},
pp. 18-31, (1958) (in Russian).
%
\bibitem{jidkov}  E.~ Jidkov et al., In ''Physics of Elementary
particles and atomic Nuclei'', vol. 4, Part. 1 (1973), JINR, Dubna (in
Russian).
%
\bibitem{BTFY} T.~Boyadjiev, M.~Todorov, P.~ Fiziev, S.~Yazadjiev,
                         {\em  Mathematical Modeling of Boson-Fermion Stars in the Generalized
                          Scalar-Tensor Theories of Gravity}, E-print:  math.SC/9911118
%
\bibitem{KMS} F.~Kusmartsev, E.~Mielke, F.~Shunk,
                         Phys. Rev.  D {\bf 43}, 3895 (1991)

\end{thebibliography}
\end{document}